\documentclass[twocolumn,showpacs,showkeys,preprintnumbers,
amsmath,amssymb,aps,floatfix,pre]{revtex4}
 
\usepackage{graphicx}
\usepackage{amsmath}
\usepackage{multirow}
\usepackage{times}
\usepackage{amssymb}
\usepackage{dcolumn}%
\usepackage{bm}%
\usepackage{hyperref}
\usepackage{ctable}
\usepackage{tabularx}

\hypersetup
{
	colorlinks=true,
	linkcolor=blue,
	citecolor=blue
}

\newcommand{\Cu}{$\mbox{Cu}$ }
\newcommand{\Zr}{$\mbox{Zr}$ }
\newcommand{\CuZr}{$\mbox{CuZr}$ }
\newcommand{\CuZrPerc}{$\mbox{Cu}_{50}\mbox{Zr}_{50}$ }
\newcommand{\CuZra}{$\mbox{Cu-Zr}$ }
\newcommand{\CuZrZr}{$\mbox{CuZr}_2$ }

\begin{document}

\title[]{{Dynamical, structural and chemical heterogeneities in a binary metallic glass-forming liquid}}
\author{F.\@ Puosi  }
\email{francesco.puosi@simap.grenoble-inp.fr}
\affiliation{Univ. Grenoble Alpes, CNRS, Grenoble INP \footnote{Institute of Engineering Univ. Grenoble Alpes}, SIMaP, F-38000 Grenoble, France}
\author{N.\@ Jakse  }
\affiliation{Univ. Grenoble Alpes, CNRS, Grenoble INP \footnote{Institute of Engineering Univ. Grenoble Alpes}, SIMaP, F-38000 Grenoble, France}
\author{A.\@ Pasturel  }
\email{alain.pasturel@simap.grenoble-inp.fr}
\affiliation{Univ. Grenoble Alpes, CNRS, Grenoble INP \footnote{Institute of Engineering Univ. Grenoble Alpes}, SIMaP, F-38000 Grenoble, France}

\date{\today}

\begin{abstract}
As approaching the glass transition, particle motion in liquids becomes highly heterogeneous and regions with virtually no mobility coexist with liquid-like domains. This complex dynamics is believed to be responsible for different phenomena including non-exponential relaxation and the breakdown of Stokes-Einstein relation. Understanding the relationships between dynamical heterogeneities and local structure in metallic liquids and glasses is a major scientific challenge. Here we use classical molecular dynamics simulations to study the atomic dynamics and microscopic structure of \CuZrPerc alloy in the supercooling regime. Dynamical heterogeneities are identified via an isoconfigurational analysis. As deeper supercooling is achieved a transition from isolated to clustering low mobility atoms is reported. These slow clusters, whose size grow upon cooling, are also associated to concentration fluctuations, characterized by a Zr-enriched phase, with a composition  \CuZrZr. In addition, a structural analysis of slow clusters based on Voronoi tessellation evidences an increase with respect of the bulk system of the fraction of Cu atoms having a local icosahedral order. These results are in agreement with the consolidated scenario of  the relevant role played by icosahedral order in the dynamic slowing-down in supercooled metal alloys. 

\end{abstract}

\maketitle

\section{Introduction}
\label{intro}

If a liquid is supercooled below its melting temperature, the glass transition (GT) can be attained, which is accompanied by a remarkable dynamic slowing down and a dramatic change in physical properties of the system. The understanding of the underlying microscopic origin of this slowing down represents a topic of much current research. One approach to study this phenomenon is to probe the temperature evolution of viscosity, self-diffusion constants and relaxation time (long time dynamics) from the very liquid phase down to the  glass transition temperature, $T_g$. Perhaps one the most striking results is the identification of a qualitative change of the dynamics at temperatures well above $T_g$ \cite{IwashitaPRL13}.  
As the the temperature is decreased in this range, several phenomena are observed: (1) the Arrhenius-to-non-Arrhenius transition of transport coefficients and relaxation \cite{DebeStilli2001,Angell91,Royall_PhysRep15}, (2) the breakdown of the Stokes-Einstein (SE) relation that accounts for the coupling of the shear viscosity $\eta$ with the translational diffusion \cite{Ediger00,Tarjus_JCP95,Xu_NatPhys09}, (3) the emergence of dynamic heterogeneities (DHs), i.e., the spatial distribution of mobilities which may differ of orders of magnitude in regions only a few nanometers away  \cite{Ediger00,Richert02,BerthieRev}. For metallic glass-forming liquids, two crossover temperatures $T_A$ and $T_S$ ($T_A > T_S$) are identified \cite{Hu_JAP2016,Soklaski_PhilMag16}. The Arrhenius-to-non-Arrhenius crossover in transport coefficients is found to associate to $T_A$ while $T_S$ marks both the fractional SE relation and the onset of DHs. In a different approach, Douglas {\textit {et al.}} \cite{Zhang_JCP15,DouglasLocalMod16} have shown that temperatures which characterize dynamics of supercooled metallic glass-forming liquids can be estimated from the temperature dependence of the picosecond Debye-Waller (DW) factor. However, how quantitatively correlates this fast-dynamics based analysis with that obtained from long-time dynamics still remains elusive.

A key aspect of  the solidification leading to glass formation is that it is associated only to subtle static structure changes. Indeed the static structure factor $S(q)$, measuring the spatial correlations of the particle positions, does not show any significant changes on approaching the glass transition. The absence of apparent static correlations distinguishing a glass and a liquid challenges the concept that structure should somehow underlie the associated dynamic phenomena. In many metallic glass-forming liquids, the marked slowing down in the supercooling regime is closely correlated to the presence of icosahedral motifs and their evolution with temperature \cite{ChengMa_ProgMatSci11,Hu_NatComm15}. More specifically, it has been found that the formation of icosahedral medium-range structures via the connectivity of full icosahedra plays a key role in dynamic arrest and glass formation in Cu-based liquids \cite{LadJCP12,Wu_PRB13,Hao_PRB11}. Interestingly, the fraction of full icosahedra is almost Cu-centered like in the popular Cu-Zr system which may yield concentration fluctuations \cite{Soklaski_PhilMag16}. However, how these concentration fluctuations qualitatively correlate to the heterogeneous dynamics is not yet investigated.

The purpose of this work is to use Molecular Dynamics (MD) simulations to analyze both fast and long-time dynamics in \CuZrPerc upon supercooling and to explore their structural and chemical basis. Our analysis reveals that the picosecond Debye-Waller factor determines the two crossover temperatures $T_{A}$ and $T_S $ in quantitative agreement with long-time dynamics. Furthermore, using the isoconfigurational ensemble method, we evidence that dynamical, structural and chemical heterogeneities are strongly correlated to DHs, characterized by the emergence of a transient low mobility phase with composition \CuZrZr.

The paper is organized as follows. In Sec. \ref{methods} details about the model and the numerical simulations are given. The results are presented and discussed in Sec. \ref{results} and the conclusions are summarized in Sec. \ref{conclusion}.

\section{Methods}
\label{methods}

Molecular Dynamics (MD) simulations for the \CuZr binary alloy were carried out using LAMMPS molecular dynamics software \cite{lammps}.  An embedded-atom model (EAM) potential was used to describe the interatomic interactions \cite{MendelevJAP09}.  Each simulation consists of a total number of $23328$ atoms contained in a box with periodic boundary conditions. The initial configurations were equilibrated at $2000 \,\mbox{K}$ for $5\,\mbox{ns}$ followed by a rapid quench to $500 \,\mbox{K}$ at a rate of $10^{11}\,\mbox{K/s}$. The quench was performed in the $\mbox{NPT}$ ensemble at zero pressure. During the quench run configurations at the temperatures of interest were collected and, after adequate relaxation, used as starting points for the production runs in the $\mbox{NVT}$ ensemble.  

\section{Results and discussion}
\label{results}

\subsection{Definition of characteristic temperatures}

First, we focus on the slowing down of dynamics entering  the supercooled regime as described by relaxation and transport properties. From the self-intermediate scattering function $F_s(q,t)$  we define the structural relaxation time $\tau_\alpha$ via the relation $F_s(q_{max},\tau)=1/e$ being $q_{max}$ the wave-vector corresponding to the main peak in the static structure factor. The viscosity $\eta$ is calculated by integrating the stress autocorrelation function according to Green-Kubo formalism \cite{ZwanzingARPC65}, i.e. $\eta=(V/k_BT)\int_0^{\infty} \langle P_{\alpha\beta}(t_0)P_{\alpha\beta}(t_0+t)\rangle dt$ where $V$ is the volume, $P_{\alpha\beta}$ is the off-diagonal $\alpha\beta$ component of the stress and an average over the three components $\alpha\beta=xy,xz,yz$ is considered. Self-diffusion coefficient $D$ is determined as the slope of the mean square displacement $\Delta r^2$ versus time $t$ in the long-time limit, $D=\lim_{t\rightarrow \infty} \frac{1}{6t} \Delta r^2(t)$. 

\begin{figure}[th]
\begin{center}
\includegraphics[width=0.85\columnwidth]{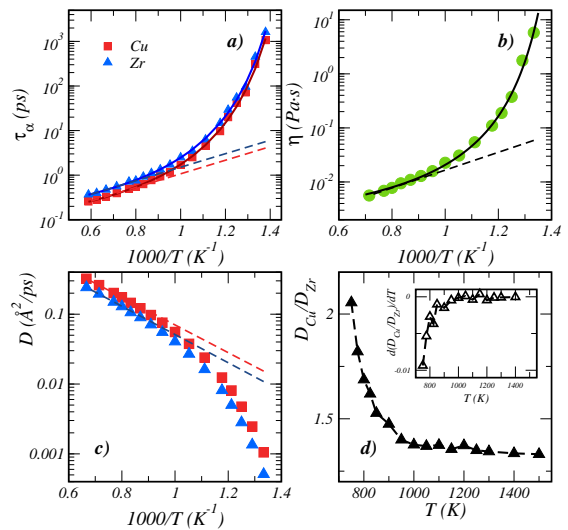}
\end{center}
\caption{Panels a), b) and c): Arrhenius plot of the structural relaxation time $\tau_\alpha$ of \Cu and \Zr atoms (panel a),  viscosity $\eta$ (panel b) and diffusion coefficient $D$  of \Cu and \Zr atoms (panel c).  Dashed lines mark  the Arrhenius behavior in the high temperature region. Full lines are fit with the Vogel-Fulcher-Tammann (VFT) laws $\tau_{\alpha}=\tau_{0} \exp\left ( A T_{0}/(T-T_{0}) \right )$   and   $\eta=\eta_{0} \exp\left ( A T_{0}/(T-T_{0}) \right )$.  Panel d): temperature dependence of the diffusion coefficient between  \Cu and \Zr atoms. }
\label{figslowing}
\end{figure}

 In Fig. \ref{figslowing} we show the temperature dependence of the structural relaxation time $\tau_\alpha$ (panel a), viscosity $\eta$ (panel b) and self diffusion coefficient $D$ (panel c).
The typical phenomenology of glass-forming liquids is observed with $\tau_\alpha$, $\eta$ and $D$ deviating from an Arrhenius temperature dependence at low temperatures. These deviations become apparent at the crossover temperature $T_A \sim 1300\,\mbox{K}$ whose estimate is in agreement with previous findings \cite{Hu_JAP2016}. Below $T_A$ the behavior of relaxation time and viscosity can be described by the Vogel-Fulcher-Tamman (VFT) equation $\tau_\alpha=\tau_0\exp [B/(T-T_0)] $ and $\eta=\eta_0\exp [B/(T-T_0)] $ respectively. The fitted valued of $T_0$ from relaxation data does not depend on the chemical species, as $T_0 = 596\,\mbox{K}$ for \Cu and $T_0 = 598\,\mbox{K}$ for \Zr. Within the uncertainty, these values are in agreement with  $T_0 = 648\,\mbox{K}$ from the fit of viscosity data.

\begin{figure}[t!]
\begin{center}
\includegraphics[width=0.85\columnwidth]{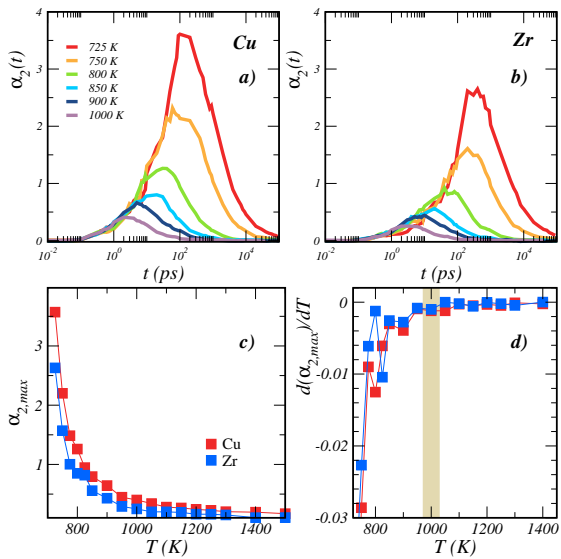}
\end{center}
\caption{ Panels a) and b): non-gaussian parameter $\alpha_2$ of \Cu (panel a) and \Zr (panel b) atoms for selected temperature values. Panel c): temperature dependence of the maximum value $\alpha_{2,max}$. Panel d):  temperature derivative $d \alpha_{2,max}/dT$ as a function of temperature. Shaded region marks the appearance of dynamical heterogeneities. }
\label{figalpha2}
\end{figure}

In the panel d)  of Fig. \ref{figslowing} the ratio of \Cu and \Zr self-diffusion coefficients is shown as a function of temperature. We note that \Cu and \Zr diffusions decouple across the temperature range considered and this decoupling is accelerated below a temperature $T\sim1000\,\mbox{K}$ as denoted by the temperature derivative in the inset of Fig. \ref{figslowing}.  The decoupling of dynamics is expected to play an important role in promoting heterogeneous dynamics in the system. To characterize dynamical heterogeneities, the non-Gaussian parameter (NGP) $\alpha_2(t)=\frac{3\langle \Delta r^4(t) \rangle }{5 \langle \Delta r^2(t) \rangle ^2}-1$ where $ \Delta r(t)$ is the displacement of a particle in a time interval $t$. Figure \ref{figalpha2} shows the NGP of \Cu (panel a) and \Zr (panel b) for selected temperatures. Deviations from the gaussian behavior first increase with time and then decay in the gaussian diffusive regime, resulting in a maximum $\alpha_{2,max}$. Non-gaussianity increases by decreasing the temperature as it's apparent from the temperature dependence of the maximum $\alpha_{2,max}$ in Figure \ref{figalpha2} (c). The change $\alpha_{2,max}$ with temperature is seen more clearly if looking at the temperature derivative $d\alpha_{2,max}/dT$ in Figure \ref{figalpha2} (d). The dynamic crossover temperature $T_s$, that marks the onset of DHs, is found to be about $1000\,\mbox{K}$: below $T_s$ we note a strong temperature dependence of $d\alpha_{2,max}/dT$ for both \Cu and \Zr atoms indicating that DHs grow with an increasing rate. The fact that the decoupling of diffusion is also accelerated below $T_s$, as mentioned earlier, is an indication of a correlation between these two phenomena.  

\begin{figure}[t!]
\begin{center}
\includegraphics[width=0.85\columnwidth]{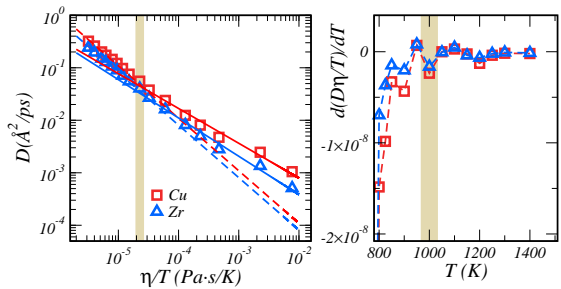}
\end{center}
\caption{ Left:  diffusion coefficient $D$ of \Cu and \Zr atoms as a function of the ratio $\eta/T$.  Dashed lines correspond to the SE relation $D\propto  (\eta/T)^{-1}$. Full lines correspond to  fractional SE relation $D\propto  (\eta/T)^{-\kappa}$. Right: breakdown of the SE relation as denoted by the temperature derivative of the product $D\eta/T$.  Shaded regions in both panels mark the onset of the breakdown of SE relation.  }
\label{figSEbreakdown}
\end{figure}

The  occurrence of DHs is indicated as the origin of the breakdown of Stokes-Einstein (SE) relation in many studies of glass-forming systems \cite{BerthieRev,LadJCP12,Puosi12SE}. SE relation provides a simple connection between the diffusion coefficient and the viscosity via $D\sim(\eta/T)^{-1}$. Figure \ref{figSEbreakdown} (left) displays the evolution of $D$ as a function of $\eta/T$. Upon cooling, we observe the breakdown of SE relation in the form of a crossover towards a fractional SE relation    $D\sim(\eta/T)^{-\zeta}$ with $\zeta=0.66$ for \Cu and $\zeta=0.72$ for \Zr. From the temperature derivative of the SE product $D\eta/T$, the breakdown is seen to occur at $T\sim1000\,\mbox{K}$, coinciding with the onset of the growth of DHs. 

\begin{figure}[th]
\begin{center}
\includegraphics[width=0.85\columnwidth]{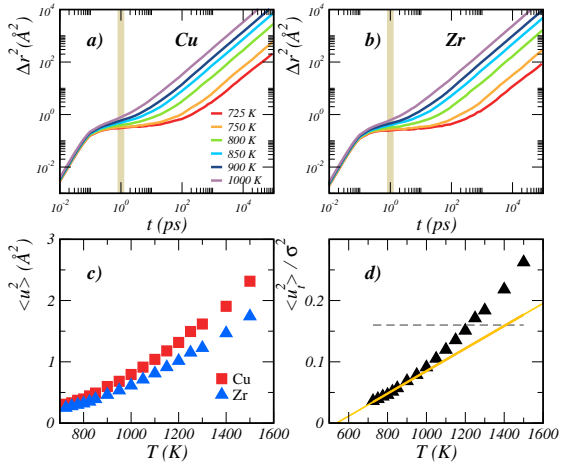}
\end{center}
\caption{ Panels a) and b): Mean square displacement of \Cu (panel a) and \Zr (panel b) atoms for selected temperature values. Vertical lines marks the time $t\sim 1\,\mbox{ps}$ at which the short time mobility $u^2$ is evaluated.  Panel c): temperature dependence of the Debye-Waller factor $\langle u^2 \rangle$ for \Cu and \Zr atoms.  Panel c): temperature dependence of the global  Debye-Waller factor $\langle u^2_t \rangle$   (average over all atoms) normalized by  the square of  the average interparticle distance $\sigma^2$. Full line corresponds to the linear fit in the low temperature range.  Dashed line indicates the value $\langle u^2 \rangle_t/\sigma^2=0.16$ which define the onset temperature $T_A$, according to a Lindemann-like criterion.}
\label{figmsd}
\end{figure}

Dynamic crossover temperatures  can also be determined by looking at short time dynamics as discussed in a recent work by Zhang \textit{et al.} \cite{Zhang_JCP15}. Here we briefly review this approach. Fast dynamics is usually described by the Debye-Waller (DW) factor $ \langle u^2  \rangle$ corresponding to the mean square vibrational amplitude of atoms. Here, we define $ \langle u^2  \rangle=\Delta r^2(t=1\,\mbox{ps})$; this definition ensures that at sufficiently low temperature $ \langle u^2  \rangle $ corresponds to the plateau value of the mean square displacement, denoting particle caging (see panel a and b of Figure \ref{figmsd}). In Figure \ref{figmsd} we show the  temperature dependence  of the DW factor for each chemical species (panel c) as well as the average over all the atoms (panel d), denoted by $ \langle u^2_t  \rangle$. First, we observe that DW extrapolates to zero at a finite temperature $T\sim540\,\mbox{K}$, in agreement with previous findings in metallic systems \cite{Zhang_JCP15} and polymers \cite{Puosi_JCP17}, and which coincides, within the experimental uncertainty, with the temperature $T_0$, estimated from the VFT fit of relaxation and viscous properties. Further, the crossover temperature $T_A$ can be estimated via an extension of the Lindemann criterion for crystal melting \cite{StilliLaViolette85}. The glass transition temperature $T_g$ is  reached when the DW attains a given fraction of the interparticle distance, i.e., $\langle u^2 (T_g) \rangle ^{1/2}\approx 0.15-0.16 \sigma $, and $T_A$ when the DW is three time this value  $\langle u^2 (T_A) \rangle ^{1/2}\approx 0.5 \sigma $. Here this numerical estimation of $T_A$ is fairly consistent with the one from $\tau$ or $\eta$. We note that a threshold for the amplitude of atomic vibrations at $T_A$ corresponding to half interparticle distance can be understood in the light of recent findings by Iwashita and coworkers \cite{IwashitaPRL13}. The authors have shown that the competitions between the elementary excitations and phonons localization determines the crossover phenomenon:  at $T_A$ local elementary excitations in atomic connectivity start to communicate with neighboring atoms through atomic vibrations as phonon mean-free path equal the distance between nearest neighbors. 

To conclude the discussion about short-time dynamics, we note that the DW factor starts to show significant deviations from the low-temperature linear behavior at about $950\,\mbox{K}$, within the uncertainty range of the crossover temperature $T_s$. Hence we interpret the superlinear increase of the DW factor as related to the onset of DHs. We point out that recently Douglas {\textit {et al.}} proposed a different interpretation as they associate the phenomenon with the critical mode-coupling temperature $T_c$ \cite{Zhang_JCP15}. For the present model a power-law fitting of the relaxation time $\tau_\alpha\sim | T-T_c|^{-\gamma}$ gives $T_c=860\,\mbox{K}$ \cite{Hu_JAP2016}. As the quality of our data does not allow us to decisively discriminate between the two proposed interpretations, further work is needed to clarify this aspect.  


To summarize, we have identified two crossover temperature characterizing the slowing down of dynamics. The Arrhenius to non-Arrhenius crossover temperature for relaxation and transport properties $T_A\sim1300\,\mbox{K}$ and temperature $T_s\sim1000\,\mbox{K}$ which characterizes different interconnected phenomena such as the appearance of dynamical heterogeneities, the decoupling of diffusion and the breakdown of SE relation. Besides, we mention the mode-coupling  temperature $T_c$ and the VFT temperature $T_{0}$; however, unlike $T_A$ and $T_s$, $T_c$  and $T_0$ are the results of fitting procedure with no clear direct connection to dynamic phenomena. 

\begin{figure}[th]
\begin{center}
\includegraphics[width=0.85\columnwidth]{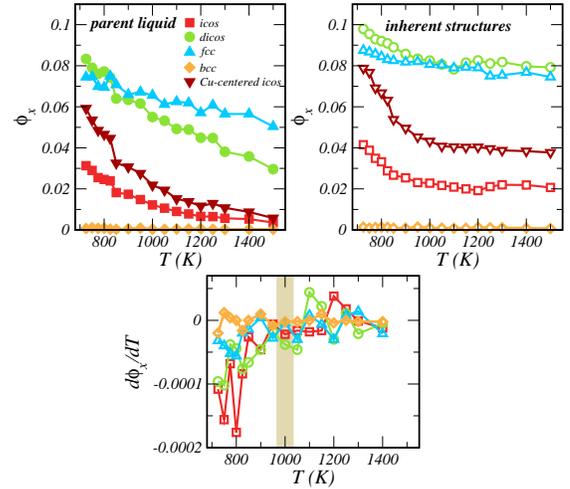}
\end{center}
\caption{ Panel a) and b): Temperature evolution of the population of Voronoi polyhedra,  Cu-centered and Zr-centered, with different symmetries (see the text for the corresponding definitions) in the parent liquid (panel a) and in the inherent structures (panel b). For comparison, data are also shown for Cu-centered full icosahedra. Panel c): temperature derivative of the polyhedra population in the inherent structures.  Shaded region marks the onset of steep increase of icosahedral-like short range order.}
\label{figvoroall}
\end{figure}

Now, we discuss the evolution of the local structure upon cooling. Short-range local ordering was investigated using the Voronoi tessellation method \cite{FinneyNature77} which consists in the decomposition of the system into a finite number of polyhedra centered at the various atomic sites. Each polyhedron is characterized by its Voronoi indices $\left \langle n_3,n_4,...,n_k,... \right \rangle $ where $n_k$ denotes the number of $k$-edged faces in the given Voronoi polyhedron (with $k$ ranging from $3$ to $10$). Different combinations of Voronoi indices corresponds to different local symmetries. Here we define the following local structures: icosahedra (icos)  for $\left \langle 0,0,12,0 \right \rangle $; distorted icosahedra (dicos) for $\left \langle 0,2,8,2 \right \rangle $, $\left \langle 0,1,10,2 \right \rangle $ and $\left \langle 1,0,9,3 \right \rangle $; face centered cubic-like (fcc) for $\left \langle 0,3,6,4 \right \rangle $, $\left \langle 0,3,6,5 \right \rangle $, $\left \langle 0,4,4,6 \right \rangle $ and $\left \langle 0,4,4,7 \right \rangle $; body centered cubic-like (bcc) for $\left \langle 0,6,0,8 \right \rangle $ \cite{Cape_JCP81,Jiang_JCP16}. In Figure \ref{figvoroall} (a,b) we plot the population of local structures as a function of temperature both in the parent liquid (panel a) and in the inherent structures (panel b) following energy minimization.
Our results are in agreement with previous simulation studies in \CuZra alloy \cite{Cheng_PRB08,LadJCP12,Soklaski_PRB13} where it has been demonstrated that icosahedral short-range order (ISRO) is correlated with dynamical slowing down: the population of full icosahedra is found to exhibit the most significant increase as approaching the glass transition. Yet, full icosahedra consist exclusively of Cu-centered polyhedra, with virtually no contribution from Zr-centered structures.  Remarkably, the increase of ISRO compares with the onset of DHs as both occurs below the crossover temperature $T_s\sim1000\,\mbox{K}$(see panel d of  Fig. \ref{figvoroall}). Conversely, the Voronoi analysis does not indicate any structural change close to the crossover temperature $T_A$.

\subsection{Analysis of heterogeneities}

In this section we  investigate the evolution of dynamical, structural and chemical heterogeneities while approaching the glass transition. With this aim, we performed simulations in the isoconfigurational  (IC) ensemble \cite{Harrowell06}. Here $100$ NVT trajectories were simulated starting from the same initial configuration but with different momenta, sampled randomly from a Boltzmann distribution. Typically, each simulation lasts for a time at least of the order of $\tau_\alpha$ to allow DHs to evolve.  Atomic mobility is evaluated via the IC averaged mean square displacement, $\left \langle  \Delta r^2_{i,\alpha}  (\Delta t) \right \rangle_{ic}= \left \langle \left | \mathbf{r}_{i,\alpha}(\Delta t) -  \mathbf{r}_{i,\alpha}(0) \right |^2 \right \rangle_{ic}$ where $\mathbf{r}_{i,\alpha}(0)$ and $\mathbf{r}_{i,\alpha}(\Delta t)$ are the positions of the $i\mbox{th}$ atoms at the beginning and after a time $\Delta t$ in the $\alpha\mbox{th}$ trajectories and the average $\left \langle ... \right \rangle_{ic}$ is taken over all the IC trajectories. Atom propensity for displacement is usually defined as $\left \langle  \Delta r^2_{i,\alpha}  (\Delta t) \right \rangle_{ic}$  on a time interval $\tau_\alpha$ or $1.5\times\tau_\alpha$. As we are interested to follow the temporal evolution of the DHs, in the following we change this time interval $\Delta t$, going from short times, corresponding to atom vibrations, up to structural relaxation. 

\begin{figure*}[t!]
\begin{center}
\includegraphics[width=0.95\textwidth]{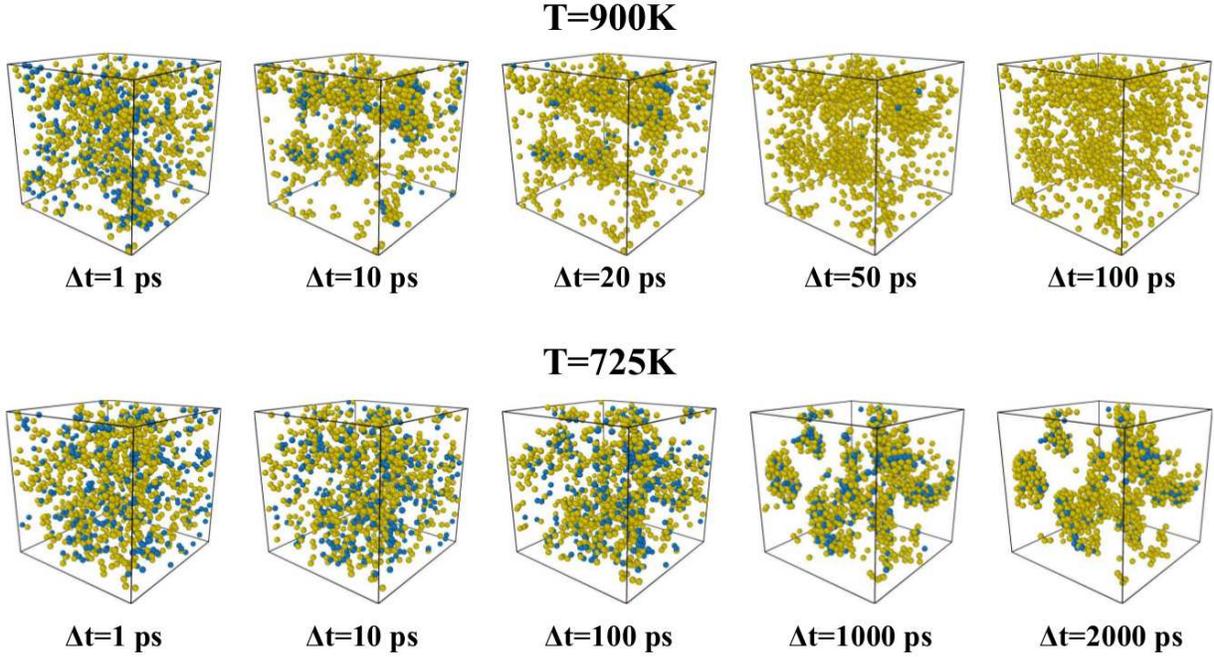}
\end{center}
\caption{  Typical configurations of low mobility (LM) atoms for different values of the time interval $\Delta t$ (see the text for definitions) and for two selected temperatures, $T=900\,\mbox{K}$ (top) and  $T=725\,\mbox{K}$ (bottom). Blue corresponds to a \Cu atom and yellow to a \Zr atom.  }
\label{figLMconfig}
\end{figure*}

Now we focus on the less mobile atoms. For each timescale of interest, we define a set of low mobility (LM) atoms as the set of $5\%$ atoms having smaller IC averaged MSD over the relevant timescale. In Figure \ref{figLMconfig}  we show the typical configurations of LM atoms for different values of the time interval $\Delta t$ and two selected temperatures. From a visual inspection we note that at the relatively high temperature, $T=900\,\mbox{K}$, the composition of the LM set of atoms evolves with the observation timescale, going from a balance of \Cu and \Zr atoms at short times to a preponderance of \Zr atoms at long times. This is indicative of a connection between compositional inhomogeneities, or chemical heterogeneities, and dynamical heterogeneities. At the lower temperature, $T=725\,\mbox{K}$, the evolution of chemical heterogeneities is also observed but the most remarkable effect is a structural evolution of LM atoms. At short observation time LM atoms are sparse in the system whereas at longer time they prefer to organize into compact clusters. 

\begin{figure}[t!]
\begin{center}
\includegraphics[width=0.75\columnwidth]{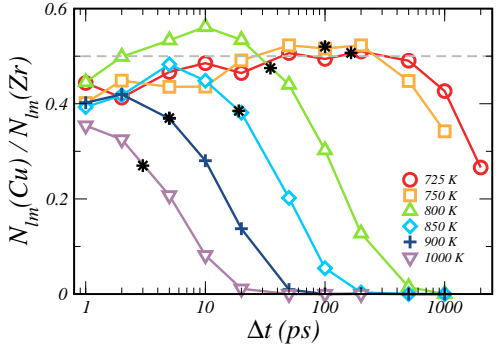}
\end{center}
\caption{ Open triangles: temporal evolution of the composition for the set of LM atoms as denoted by the ratio between  \Cu and \Zr atoms. Data are shown for different temperatures. Stars: for each temperature, composition on the time scale $t^*$ corresponding to the maximum of the non-gaussian parameter $\alpha_2$ for \Cu atoms.  }
\label{figcompo}
\end{figure}

In order to deepen the discussion concerning composition inhomogeneities, we plot in Figure \ref{figcompo} the time dependence of the composition of LM group, as denoted by the ratio between \Cu and \Zr atoms, $N_s(\mbox{Cu})/N_s(\mbox{Zr})$, for a temperature range corresponding to increasing DHs. 
As mentioned above composition shows a clear dependence on the observation time scale $\Delta t$. At short times, the composition, which is different from the macroscopic one, seems not to depend on the temperature. At long times, beyond structural relaxation, LM atoms consist virtually of only \Zr atoms. This effect is clearly due to \Zr having a lower diffusivity with respect to \Cu. At intermediate times, the ratio develops a maximum for times of the order of $t^*$, with a value $N_s(\mbox{Cu})/N_s(\mbox{Zr}) \sim 0.5$ which again seems not to depend on temperature. Upon increasing the supercooling, the maximum converts to a plateau, whose extension grows in time. This  signals the appearance of a transient low mobility phase with composition \CuZrZr at the maximum of heterogeneity in the system. We point out that the proposed scenario, and in particular the presence of the transient \CuZrZr phase, is stable with respect to the definition of the LM set of atoms (not shown here).  

\begin{figure}[t!]
\begin{center}
\includegraphics[width=0.95\columnwidth]{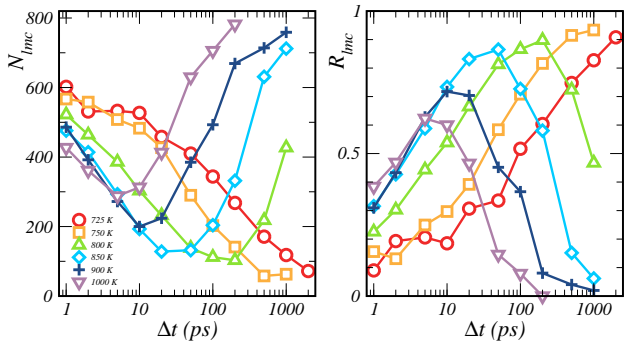}
\end{center}
\caption{  Left panel: number of clusters $N_{lmc}$ composed of LM atoms as a function of the time lag $\Delta t$ for different temperatures. Right panel: temporal evolution of the population of LM atoms $R_{lmc}$ belonging to significant clusters (clusters with more than $10$ members), normalized by the total number of LM atoms.  }
\label{figclusters1}
\end{figure}

The connections between structural and dynamical heterogeneities are examined by looking at the spatial organization of LM atoms. To this purpose, we performed a cluster analysis of the LM set. A cluster is defined as a set of connected particles, each of which is within a cutoff distance of one or more other particles from the same cluster; we set the cutoff as the first minimum in the radial pair distribution function. Figure \ref{figclusters1} shows, for selected temperature values, the time evolution of the number of LM cluster, $N_{lmc}$, and the fraction of LM atoms participating to significant clusters $R_{lmc}$ (defined as clusters with at least $10$ atoms). 
The decrease of $N_{lmc}$, as well as the increase of $R_{lmc}$, with increasing  observation time signals the tendency of LM atoms to form clusters, as already noted by visual inspection. Clustering becomes stronger as the temperature decreases: at the lowest investigated temperatures about $90\%$ of LM atoms are involved in clusters. At long times approaching the diffusion regime, as \Zr starts to play the dominant role,  a change back to isolated LM atoms is observed. 

\begin{figure}[t!]
\begin{center}
\includegraphics[width=0.95\columnwidth]{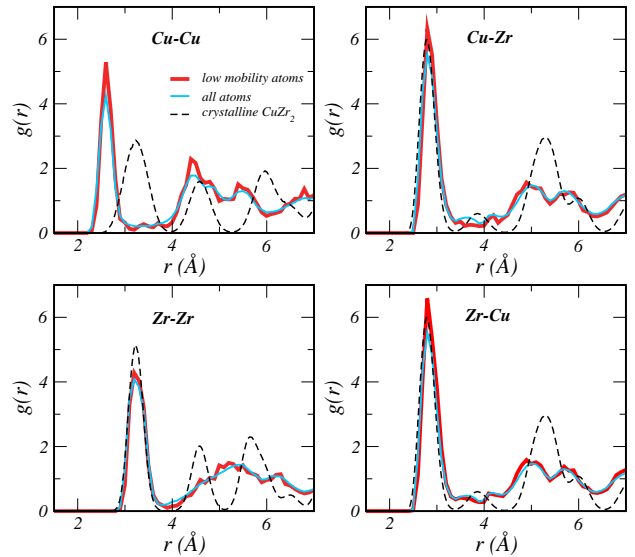}
\end{center}
\caption{Partial radial pair distribution function $g(r)$.  Thick full lines: $g(r)$ for a central atom belonging to the transient LM  phase at $T=750\,\mbox{K}$. Thin full lines:  global $g(r)$  $T=750\,\mbox{K}$. Dashed lines: $g(r)$ for $\mbox{CuZr}_2$ crystal at $T=750\,\mbox{K}$.   }
\label{figgdr}
\end{figure}

Before going any further in the structural characterization of LM atoms, we come back to the observation of a transient low mobility phase with composition \CuZrZr. To understand what is the role of this phase, one has to consider the phase diagram of the \CuZra alloy. Indeed, for the composition $\mbox{Cu}_{50}\mbox{Zr}_{50}$ the compound \CuZrZr is known to form as the primary crystalline phase from the undercooled liquid below $800\,\mbox{K}$ \cite{Abe_AM06}. This similarity raises the question of whether the low mobility \CuZrZr is somehow related to the crystalline  phase. To test this hypothesis we have performed MD simulations of the \CuZrZr crystal at finite temperature, obtained  starting from a crystalline configuration at zero temperature which was heated to temperature of interest in the NVT ensemble.   Local atomic environment in \CuZrZr crystal is known from literature data \cite{CuZr_crystal}. Here \Cu sites have coordination $14$, resulting from $4$ \Cu atoms with equilibrium distance $d_1=3.218\,\mbox{\AA}$ and $10$ \Zr atoms at distances $d_2=2.848\,\mbox{\AA}$ (8 atoms) and $d_3=3.877\,\mbox{\AA}$ (2 atoms). \Zr sites have coordination $14$: $5$ \Cu atoms with equilibrium distance $d_2=2.848\,\mbox{\AA}$ (4 atoms) and $d_3=3.877\,\mbox{\AA}$ (1 atom) and $9$ \Zr atoms at distances $d_4=3.140\,\mbox{\AA}$ (4 atoms), $d_1=3.218\,\mbox{\AA}$ (4 atoms) and $d_3=3.877\,\mbox{\AA}$ (1 atom). In Figure \ref{figgdr} we show the partial pair distribution function $g(r)$ for a central atom belonging to the transient LM phase at $T=750\,\mbox{K}$ and the corresponding $g(r)$ of the crystal at the same temperature. As a reference, we also plot the $g(r)$ for all the atoms in the system. First, for a consistency check, we note that simulated \CuZrZr crystal is in agreement with the reported literature data. Then we note that the pair distribution function for LM atoms doesn't show any significant difference with respect to the full system averaged $g(r)$.  This emphasizes that any change in local order of the LM \CuZrZr phase cannot be detected from $g(r)$. Indeed, on the one hand the position and height of the first peak in the $g(r)$ for LM \Zr coincide with those of the crystal for both $\mbox{Zr}-\mbox{Zr}$ and $\mbox{Zr}-\mbox{Cu}$ pairs (bottom panels in Fig. \ref{figgdr}), on the other hand the $g(r)$ for LM \Cu atoms shows strong difference with the corresponding crystalline one (top panels in Fig. \ref{figgdr}). The mismatch in the position of the first peak is a signature of a considerable difference in local density between the LM and crystalline \CuZrZr phases. This comparison on the local atomic environment seems to indicate that no  direct connection exists between the observed low mobility regions and the nucleation of the primary crystalline phase.   

\begin{figure}[t!]
\begin{center}
\includegraphics[width=0.95\columnwidth]{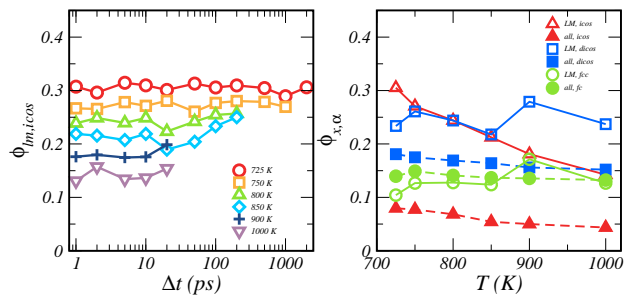}
\end{center}
\caption{ Left:  fraction of  \Cu LM atoms with icosahedral-like short-range order as a function of the observation time scale $\Delta t$. Data are shown for selected temperature values. Right: temperature dependence of the  \Cu fraction in the LM set (open symbols) and in the all system (full symbols) with icosahedral (icos), distorted-icosahedral (dicos) and fcc-like (fcc) short range order.  }
\label{figvorolm}
\end{figure}

To complete the structural characterization of DHs, we revert to the Voronoi analysis of LM atoms. According to the previous findings, we focus on icosahedral-like SRO for \Cu centered polyhedra, which is known to play the key role. In Figure \ref{figvorolm} (left) the fraction of \Cu LM atoms with icosahedral-like short-range order, $\phi_{lm,icos}$,  is shown as a function of the observation time scale $\Delta t$. 
For all the considered temperatures, the population of LM icosahedra displays a flat dependence on $\Delta t$. Only for high temperatures a decrease is observed at long times (not shown), due to the loss in the LM set of \Cu atoms. In addition, the mean value of $\phi_{lm,icos}$ increases upon cooling. This can be seen more clearly in Figure \ref{figvorolm} (right) where the temperature dependence of $\phi_{lm,icos}$ is shown, together with the total fraction of \Cu atoms with ISRO in the system. Interestingly, in the considered temperature range the LM set is $3-4$ times reacher in full icosahedra with respect to the whole system.  On the contrary, other types of Voronoi polyhedra, such as distorted icosahedra and fcc-like, show for LM atoms both no significant temperature dependence and a weaker increase if compared to the whole system.

\begin{figure}[t!]
\begin{center}
\includegraphics[width=0.48\columnwidth]{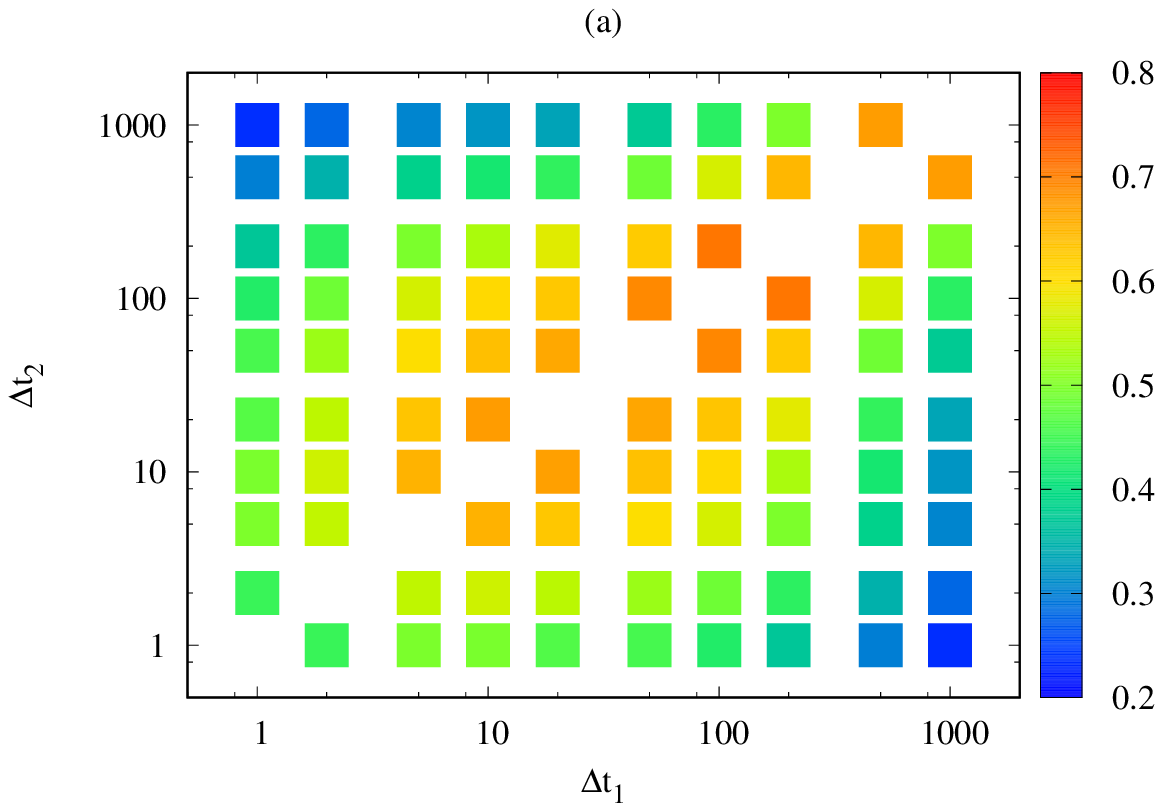}
\includegraphics[width=0.48\columnwidth]{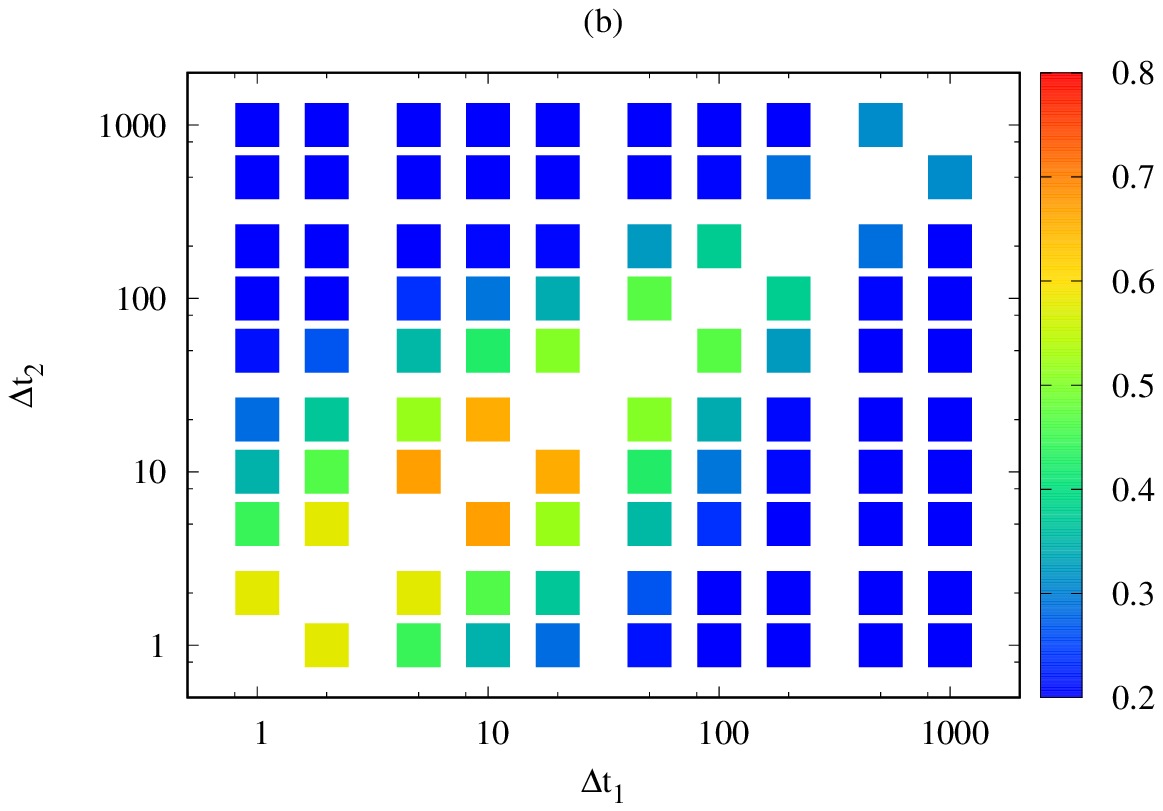}
\includegraphics[width=0.48\columnwidth]{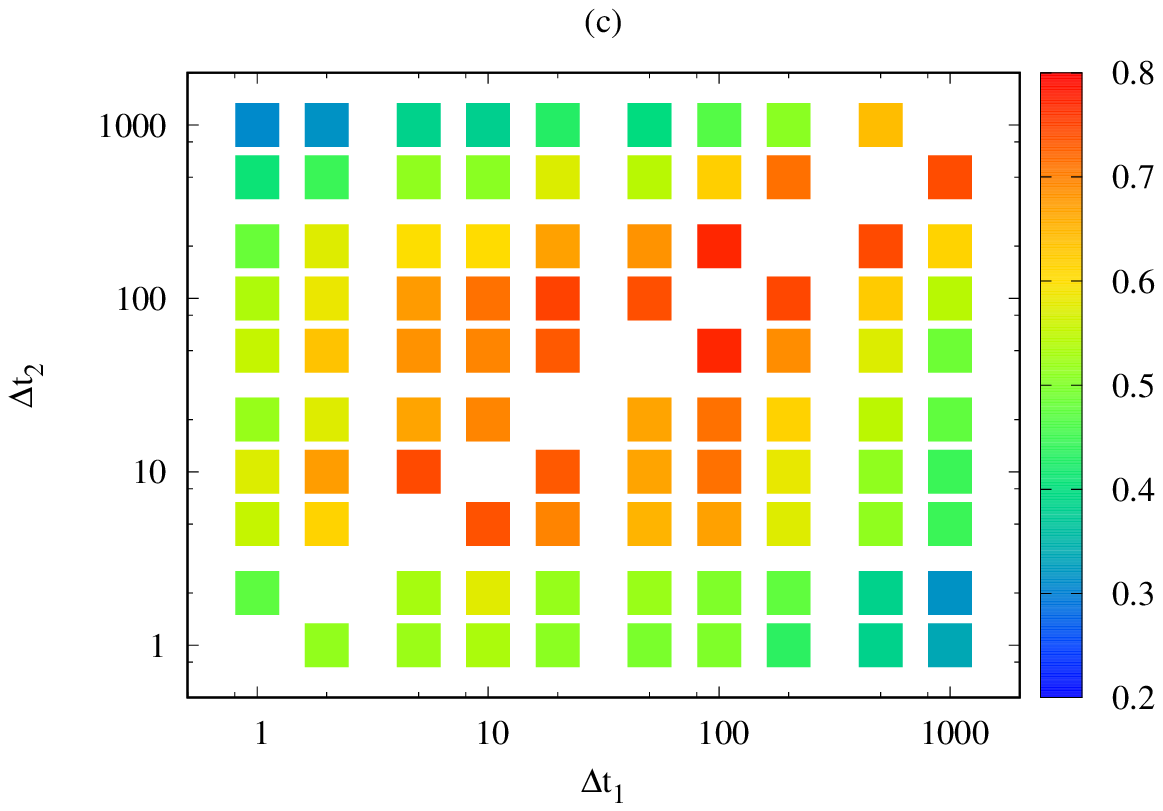}
\includegraphics[width=0.48\columnwidth]{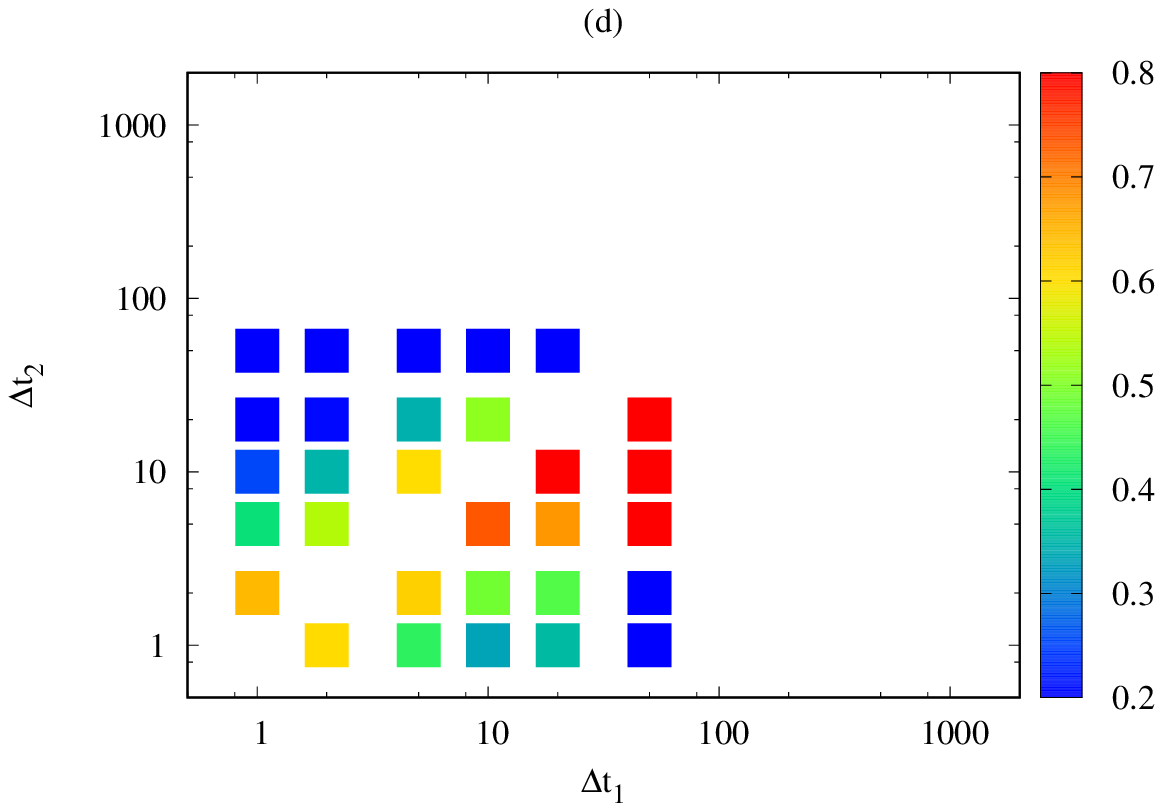}
\end{center}
\caption{ Panels (a) and (b): Two-time correlation function of the LM atoms $\Psi_{lm}(\Delta t_1, \Delta t_2)$, denoting the probability that a LM atom over the timescale $\Delta t_1$ is also  a LM atom on the time interval $\Delta t_2$. Data are shown for $T=725\,\mbox{K}$ (panel a) and $T=900\,\mbox{K}$ (panel b). Panels (c) and (d): $\Psi_{lm}(\Delta t_1, \Delta t_2)$, restricted to the atoms  with icosahedral short-range order on the observation time  $\Delta t_1$. Data are shown for $T=725\,\mbox{K}$ (panel c) and $T=900\,\mbox{K}$ (panel d).}
\label{figcorr}
\end{figure}

Finally we address the issue of the stability of DHs. In this spirit, we define the two-time low-mobility correlation function $\Psi_{lm}(\Delta t_1, \Delta t_2)$ which quantifies the probability that a LM atom over the time interval $\Delta t_1$ is also  a LM atom on the time interval $\Delta t_2$. In Figure \ref{figcorr} (a,b) we compare $\Psi_{lm}(\Delta t_1, \Delta t_2)$ for two selected temperatures $T=725\,\mbox{K}$ (a) and $T=900\,\mbox{K}$ (b), corresponding to strong and moderate supercooling respectively. For short observation times $\Delta t_1$, corresponding to fast vibrational dynamics,  DHs are rather unstable as the value of $\Psi_{lm}$ drops below $0.5$ within few picoseconds. As one could expect, for both temperatures, the most persistent DHs are found when   $\Delta t_1$ is of the order of $t^*$, the time of the maximum in the NGP. Here, if we fix $\Delta t_1\sim t^*$, we note that the effect of decreasing the temperature is to slow-down the decrease of $\Delta t_1$, i.e. to increase the stability of LM atoms. 
To go further, in Figure \ref{figcorr} (c,d) we also plot the correlation function $\Psi_{lm}(\Delta t_1, \Delta t_2)$ restricted to LM atoms with ISRO on the time interval $\Delta t_1$. For the lowest temperature $T=725\,\mbox{K}$ the presence of ISRO results clearly in an increased stability of LM atoms, in particular in the region $\Delta t_1\sim t^*$. This is also seen in the higher temperature, at the onset of DHs, even though here the competing phenomenon of loosing LM atoms with icosahedral environment as increasing $\Delta t_1$ makes the effect less apparent.

\section{Conclusion}
\label{conclusion}

The present MD study aims at clarifying the relationship between dynamics, chemistry and structure in supercooled liquids by investigating the interplay between structural, chemical and dynamical heterogeneities in a model of metallic glass-forming liquid, namely \CuZrPerc alloy. As a preliminary step, we have analyzed the temperature evolution of fast dynamics and slow relaxation upon supercooling and identified two crossover temperature $T_A$ and $T_s$, corresponding to the Arrhenius-to-non-Arrhenius crossover and the onset of dynamical heterogeneities respectively.  Further, by changing the observation timescale,  we were able to expose both a structural and a chemical evolution of dynamical heterogeneities. Below $T_s$ low mobility atoms are organized into clusters, whose size increases upon cooling and having a Zr-enriched composition with respect to whole system. For observation time corresponding to the maximum of DHs, we report a transient LM phase with a composition \CuZrZr which is reminiscent of the primary crystalline phase below 800K. However, a local structural analysis of this transient phase shows no signature of crystalline order but rather an increase of icosahedral short-range order (ISRO).

Our results support the consolidated scenario of a correlation between an increase of ISRO and the slowing down of dynamics as approaching the glass transition. Crystallization, which is definitely observed under experimental condition, is not attainable in our simulations due to limited accessible timescales. This precludes the possibility to investigate the competition and interplay of crystalline and amorphous ``order'' and their relationships with dynamics. To address this topic, which in our opinion deserves further attention, in future works we plan  to focus on different compositions for  \CuZr or on different models of metallic systems, for which crystallization is numerically observed  \cite{Wu_JCP13,Jiang_JCP16,Levchenko_ActaMat11}. 

\begin{acknowledgments}
\end{acknowledgments}

This work was granted access to the HPC resources of IDRIS under the allocation 2017-A0020910083 made by GENCI. Some the computations presented in this paper were performed using the Froggy platform of the CIMENT infrastructure (https://ciment.ujf-grenoble.fr), which is supported by the Rh\^{o}ne-Alpes region and the Equip@Meso project (reference ANR-10-EQPX-29-01) of the programme Investissements d'Avenir supervised by the Agence Nationale pour la Recherche.


\begin{thebibliography}{34}
\expandafter\ifx\csname natexlab\endcsname\relax\def\natexlab#1{#1}\fi
\expandafter\ifx\csname bibnamefont\endcsname\relax
  \def\bibnamefont#1{#1}\fi
\expandafter\ifx\csname bibfnamefont\endcsname\relax
  \def\bibfnamefont#1{#1}\fi
\expandafter\ifx\csname citenamefont\endcsname\relax
  \def\citenamefont#1{#1}\fi
\expandafter\ifx\csname url\endcsname\relax
  \def\url#1{\texttt{#1}}\fi
\expandafter\ifx\csname urlprefix\endcsname\relax\def\urlprefix{URL }\fi
\providecommand{\bibinfo}[2]{#2}
\providecommand{\eprint}[2][]{\url{#2}}

\bibitem[{\citenamefont{Iwashita et~al.}(2013)\citenamefont{Iwashita,
  Nicholson, and Egami}}]{IwashitaPRL13}
\bibinfo{author}{\bibfnamefont{T.}~\bibnamefont{Iwashita}},
  \bibinfo{author}{\bibfnamefont{D.~M.} \bibnamefont{Nicholson}},
  \bibnamefont{and} \bibinfo{author}{\bibfnamefont{T.}~\bibnamefont{Egami}},
  \bibinfo{journal}{Phys. Rev. Lett.} \textbf{\bibinfo{volume}{110}},
  \bibinfo{pages}{205504} (\bibinfo{year}{2013}).

\bibitem[{\citenamefont{Debenedetti and Stillinger}(2001)}]{DebeStilli2001}
\bibinfo{author}{\bibfnamefont{P.~G.} \bibnamefont{Debenedetti}}
  \bibnamefont{and} \bibinfo{author}{\bibfnamefont{F.~H.}
  \bibnamefont{Stillinger}}, \bibinfo{journal}{Nature}
  \textbf{\bibinfo{volume}{410}}, \bibinfo{pages}{259} (\bibinfo{year}{2001}).

\bibitem[{\citenamefont{Angell}(1991)}]{Angell91}
\bibinfo{author}{\bibfnamefont{C.}~\bibnamefont{Angell}},
  \bibinfo{journal}{J.Non-Crystalline Sol.} \textbf{\bibinfo{volume}{131-133}},
  \bibinfo{pages}{13} (\bibinfo{year}{1991}).

\bibitem[{\citenamefont{Royall and Williams}(2015)}]{Royall_PhysRep15}
\bibinfo{author}{\bibfnamefont{C.~P.} \bibnamefont{Royall}} \bibnamefont{and}
  \bibinfo{author}{\bibfnamefont{S.~R.} \bibnamefont{Williams}},
  \bibinfo{journal}{Physics Reports} \textbf{\bibinfo{volume}{560}},
  \bibinfo{pages}{1 } (\bibinfo{year}{2015}).

\bibitem[{\citenamefont{Ediger}(2000)}]{Ediger00}
\bibinfo{author}{\bibfnamefont{M.~D.} \bibnamefont{Ediger}},
  \bibinfo{journal}{Annu. Rev. Phys. Chem.} \textbf{\bibinfo{volume}{51}},
  \bibinfo{pages}{99} (\bibinfo{year}{2000}).

\bibitem[{\citenamefont{Tarjus and Kivelson}(1995)}]{Tarjus_JCP95}
\bibinfo{author}{\bibfnamefont{G.}~\bibnamefont{Tarjus}} \bibnamefont{and}
  \bibinfo{author}{\bibfnamefont{D.}~\bibnamefont{Kivelson}},
  \bibinfo{journal}{The Journal of Chemical Physics}
  \textbf{\bibinfo{volume}{103}}, \bibinfo{pages}{3071} (\bibinfo{year}{1995}).

\bibitem[{\citenamefont{Xu et~al.}(2009)\citenamefont{Xu, Mallamace, Yan,
  Starr, Buldyrev, and Eugene~Stanley}}]{Xu_NatPhys09}
\bibinfo{author}{\bibfnamefont{L.}~\bibnamefont{Xu}},
  \bibinfo{author}{\bibfnamefont{F.}~\bibnamefont{Mallamace}},
  \bibinfo{author}{\bibfnamefont{Z.}~\bibnamefont{Yan}},
  \bibinfo{author}{\bibfnamefont{F.~W.} \bibnamefont{Starr}},
  \bibinfo{author}{\bibfnamefont{S.~V.} \bibnamefont{Buldyrev}},
  \bibnamefont{and}
  \bibinfo{author}{\bibfnamefont{H.}~\bibnamefont{Eugene~Stanley}},
  \bibinfo{journal}{Nat Phys} \textbf{\bibinfo{volume}{5}},
  \bibinfo{pages}{565} (\bibinfo{year}{2009}).

\bibitem[{\citenamefont{Richert}(2002)}]{Richert02}
\bibinfo{author}{\bibfnamefont{R.}~\bibnamefont{Richert}}, \bibinfo{journal}{J.
  Phys.: Condens. Matter} \textbf{\bibinfo{volume}{14}}, \bibinfo{pages}{R703}
  (\bibinfo{year}{2002}).

\bibitem[{\citenamefont{Berthier and Biroli}(2011)}]{BerthieRev}
\bibinfo{author}{\bibfnamefont{L.}~\bibnamefont{Berthier}} \bibnamefont{and}
  \bibinfo{author}{\bibfnamefont{G.}~\bibnamefont{Biroli}},
  \bibinfo{journal}{Rev. Mod. Phys.} \textbf{\bibinfo{volume}{83}},
  \bibinfo{pages}{587} (\bibinfo{year}{2011}).

\bibitem[{\citenamefont{Hu et~al.}(2016)\citenamefont{Hu, Li, Li, Bai, and
  Wang}}]{Hu_JAP2016}
\bibinfo{author}{\bibfnamefont{Y.~C.} \bibnamefont{Hu}},
  \bibinfo{author}{\bibfnamefont{F.~X.} \bibnamefont{Li}},
  \bibinfo{author}{\bibfnamefont{M.~Z.} \bibnamefont{Li}},
  \bibinfo{author}{\bibfnamefont{H.~Y.} \bibnamefont{Bai}}, \bibnamefont{and}
  \bibinfo{author}{\bibfnamefont{W.~H.} \bibnamefont{Wang}},
  \bibinfo{journal}{Journal of Applied Physics} \textbf{\bibinfo{volume}{119}},
  \bibinfo{pages}{205108} (\bibinfo{year}{2016}).

\bibitem[{\citenamefont{Soklaski et~al.}(2016)\citenamefont{Soklaski, Tran,
  Nussinov, Kelton, and Yang}}]{Soklaski_PhilMag16}
\bibinfo{author}{\bibfnamefont{R.}~\bibnamefont{Soklaski}},
  \bibinfo{author}{\bibfnamefont{V.}~\bibnamefont{Tran}},
  \bibinfo{author}{\bibfnamefont{Z.}~\bibnamefont{Nussinov}},
  \bibinfo{author}{\bibfnamefont{K.~F.} \bibnamefont{Kelton}},
  \bibnamefont{and} \bibinfo{author}{\bibfnamefont{L.}~\bibnamefont{Yang}},
  \bibinfo{journal}{Philosophical Magazine} \textbf{\bibinfo{volume}{96}},
  \bibinfo{pages}{1212} (\bibinfo{year}{2016}).

\bibitem[{\citenamefont{Zhang et~al.}(2015)\citenamefont{Zhang, Zhong, Douglas,
  Wang, Cao, Zhang, and Jiang}}]{Zhang_JCP15}
\bibinfo{author}{\bibfnamefont{H.}~\bibnamefont{Zhang}},
  \bibinfo{author}{\bibfnamefont{C.}~\bibnamefont{Zhong}},
  \bibinfo{author}{\bibfnamefont{J.~F.} \bibnamefont{Douglas}},
  \bibinfo{author}{\bibfnamefont{X.}~\bibnamefont{Wang}},
  \bibinfo{author}{\bibfnamefont{Q.}~\bibnamefont{Cao}},
  \bibinfo{author}{\bibfnamefont{D.}~\bibnamefont{Zhang}}, \bibnamefont{and}
  \bibinfo{author}{\bibfnamefont{J.-Z.} \bibnamefont{Jiang}},
  \bibinfo{journal}{The Journal of Chemical Physics}
  \textbf{\bibinfo{volume}{142}}, \bibinfo{pages}{164506}
  (\bibinfo{year}{2015}).

\bibitem[{\citenamefont{Douglas et~al.}(2016)\citenamefont{Douglas,
  {Pazmi{\~{n}}o Betancourt}, Tong, and Zhang}}]{DouglasLocalMod16}
\bibinfo{author}{\bibfnamefont{J.~F.} \bibnamefont{Douglas}},
  \bibinfo{author}{\bibfnamefont{B.~A.} \bibnamefont{{Pazmi{\~{n}}o
  Betancourt}}}, \bibinfo{author}{\bibfnamefont{X.}~\bibnamefont{Tong}},
  \bibnamefont{and} \bibinfo{author}{\bibfnamefont{H.}~\bibnamefont{Zhang}},
  \bibinfo{journal}{J. Stat. Mech.: Theory Exp.} p. \bibinfo{pages}{054048}
  (\bibinfo{year}{2016}).

\bibitem[{\citenamefont{Cheng and Ma}(2011)}]{ChengMa_ProgMatSci11}
\bibinfo{author}{\bibfnamefont{Y.}~\bibnamefont{Cheng}} \bibnamefont{and}
  \bibinfo{author}{\bibfnamefont{E.}~\bibnamefont{Ma}},
  \bibinfo{journal}{Progress in Materials Science}
  \textbf{\bibinfo{volume}{56}}, \bibinfo{pages}{379 } (\bibinfo{year}{2011}).

\bibitem[{\citenamefont{Hu et~al.}(2015)\citenamefont{Hu, Li, Li, Bai, and
  Wang}}]{Hu_NatComm15}
\bibinfo{author}{\bibfnamefont{Y.~C.} \bibnamefont{Hu}},
  \bibinfo{author}{\bibfnamefont{F.~X.} \bibnamefont{Li}},
  \bibinfo{author}{\bibfnamefont{M.~Z.} \bibnamefont{Li}},
  \bibinfo{author}{\bibfnamefont{H.~Y.} \bibnamefont{Bai}}, \bibnamefont{and}
  \bibinfo{author}{\bibfnamefont{W.~H.} \bibnamefont{Wang}},
  \bibinfo{journal}{Nature Communications} \textbf{\bibinfo{volume}{6}},
  \bibinfo{pages}{8310 EP } (\bibinfo{year}{2015}).

\bibitem[{\citenamefont{Lad et~al.}(2012)\citenamefont{Lad, Jakse, and
  Pasturel}}]{LadJCP12}
\bibinfo{author}{\bibfnamefont{K.~N.} \bibnamefont{Lad}},
  \bibinfo{author}{\bibfnamefont{N.}~\bibnamefont{Jakse}}, \bibnamefont{and}
  \bibinfo{author}{\bibfnamefont{A.}~\bibnamefont{Pasturel}},
  \bibinfo{journal}{The Journal of Chemical Physics}
  \textbf{\bibinfo{volume}{136}}, \bibinfo{pages}{104509}
  (\bibinfo{year}{2012}).

\bibitem[{\citenamefont{Wu et~al.}(2013{\natexlab{a}})\citenamefont{Wu, Li,
  Wang, and Liu}}]{Wu_PRB13}
\bibinfo{author}{\bibfnamefont{Z.~W.} \bibnamefont{Wu}},
  \bibinfo{author}{\bibfnamefont{M.~Z.} \bibnamefont{Li}},
  \bibinfo{author}{\bibfnamefont{W.~H.} \bibnamefont{Wang}}, \bibnamefont{and}
  \bibinfo{author}{\bibfnamefont{K.~X.} \bibnamefont{Liu}},
  \bibinfo{journal}{Phys. Rev. B} \textbf{\bibinfo{volume}{88}},
  \bibinfo{pages}{054202} (\bibinfo{year}{2013}{\natexlab{a}}).

\bibitem[{\citenamefont{Hao et~al.}(2011)\citenamefont{Hao, Wang, Li,
  Napolitano, and Ho}}]{Hao_PRB11}
\bibinfo{author}{\bibfnamefont{S.~G.} \bibnamefont{Hao}},
  \bibinfo{author}{\bibfnamefont{C.~Z.} \bibnamefont{Wang}},
  \bibinfo{author}{\bibfnamefont{M.~Z.} \bibnamefont{Li}},
  \bibinfo{author}{\bibfnamefont{R.~E.} \bibnamefont{Napolitano}},
  \bibnamefont{and} \bibinfo{author}{\bibfnamefont{K.~M.} \bibnamefont{Ho}},
  \bibinfo{journal}{Phys. Rev. B} \textbf{\bibinfo{volume}{84}},
  \bibinfo{pages}{064203} (\bibinfo{year}{2011}).

\bibitem[{\citenamefont{Plimpton}(1995)}]{lammps}
\bibinfo{author}{\bibfnamefont{S.}~\bibnamefont{Plimpton}},
  \bibinfo{journal}{J. Comput. Phys.} \textbf{\bibinfo{volume}{117}},
  \bibinfo{pages}{1} (\bibinfo{year}{1995}).

\bibitem[{\citenamefont{Mendelev et~al.}(2009)\citenamefont{Mendelev, Kramer,
  Ott, Sordelet, Yagodin, and Popel}}]{MendelevJAP09}
\bibinfo{author}{\bibfnamefont{M.}~\bibnamefont{Mendelev}},
  \bibinfo{author}{\bibfnamefont{M.}~\bibnamefont{Kramer}},
  \bibinfo{author}{\bibfnamefont{R.}~\bibnamefont{Ott}},
  \bibinfo{author}{\bibfnamefont{D.}~\bibnamefont{Sordelet}},
  \bibinfo{author}{\bibfnamefont{D.}~\bibnamefont{Yagodin}}, \bibnamefont{and}
  \bibinfo{author}{\bibfnamefont{P.}~\bibnamefont{Popel}},
  \bibinfo{journal}{Philosophical Magazine} \textbf{\bibinfo{volume}{89}},
  \bibinfo{pages}{967} (\bibinfo{year}{2009}).

\bibitem[{\citenamefont{Zwanzig}(1965)}]{ZwanzingARPC65}
\bibinfo{author}{\bibfnamefont{R.}~\bibnamefont{Zwanzig}},
  \bibinfo{journal}{Annual Review of Physical Chemistry}
  \textbf{\bibinfo{volume}{16}}, \bibinfo{pages}{67} (\bibinfo{year}{1965}).

\bibitem[{\citenamefont{Puosi and Leporini}(2012)}]{Puosi12SE}
\bibinfo{author}{\bibfnamefont{F.}~\bibnamefont{Puosi}} \bibnamefont{and}
  \bibinfo{author}{\bibfnamefont{D.}~\bibnamefont{Leporini}},
  \bibinfo{journal}{J. Chem. Phys.} \textbf{\bibinfo{volume}{136}},
  \bibinfo{pages}{211101} (\bibinfo{year}{2012}).

\bibitem[{\citenamefont{Puosi et~al.}(2016)\citenamefont{Puosi, Chulkin,
  Bernini, Capaccioli, and Leporini}}]{Puosi_JCP17}
\bibinfo{author}{\bibfnamefont{F.}~\bibnamefont{Puosi}},
  \bibinfo{author}{\bibfnamefont{O.}~\bibnamefont{Chulkin}},
  \bibinfo{author}{\bibfnamefont{S.}~\bibnamefont{Bernini}},
  \bibinfo{author}{\bibfnamefont{S.}~\bibnamefont{Capaccioli}},
  \bibnamefont{and} \bibinfo{author}{\bibfnamefont{D.}~\bibnamefont{Leporini}},
  \bibinfo{journal}{The Journal of Chemical Physics}
  \textbf{\bibinfo{volume}{145}}, \bibinfo{pages}{234904}
  (\bibinfo{year}{2016}).

\bibitem[{\citenamefont{LaViolette and Stillinger}(1985)}]{StilliLaViolette85}
\bibinfo{author}{\bibfnamefont{R.~A.} \bibnamefont{LaViolette}}
  \bibnamefont{and} \bibinfo{author}{\bibfnamefont{F.~H.}
  \bibnamefont{Stillinger}}, \bibinfo{journal}{J. Chem. Phys.}
  \textbf{\bibinfo{volume}{83}}, \bibinfo{pages}{4079} (\bibinfo{year}{1985}).

\bibitem[{\citenamefont{Finney}(1977)}]{FinneyNature77}
\bibinfo{author}{\bibfnamefont{J.~L.} \bibnamefont{Finney}},
  \bibinfo{journal}{Nature} \textbf{\bibinfo{volume}{266}},
  \bibinfo{pages}{309} (\bibinfo{year}{1977}).

\bibitem[{\citenamefont{Cape et~al.}(1981)\citenamefont{Cape, Finney, and
  Woodcock}}]{Cape_JCP81}
\bibinfo{author}{\bibfnamefont{J.~N.} \bibnamefont{Cape}},
  \bibinfo{author}{\bibfnamefont{J.~L.} \bibnamefont{Finney}},
  \bibnamefont{and} \bibinfo{author}{\bibfnamefont{L.~V.}
  \bibnamefont{Woodcock}}, \bibinfo{journal}{The Journal of Chemical Physics}
  \textbf{\bibinfo{volume}{75}}, \bibinfo{pages}{2366} (\bibinfo{year}{1981}).

\bibitem[{\citenamefont{Jiang et~al.}(2016)\citenamefont{Jiang, Wu, and
  Li}}]{Jiang_JCP16}
\bibinfo{author}{\bibfnamefont{S.~Q.} \bibnamefont{Jiang}},
  \bibinfo{author}{\bibfnamefont{Z.~W.} \bibnamefont{Wu}}, \bibnamefont{and}
  \bibinfo{author}{\bibfnamefont{M.~Z.} \bibnamefont{Li}},
  \bibinfo{journal}{The Journal of Chemical Physics}
  \textbf{\bibinfo{volume}{144}}, \bibinfo{pages}{154502}
  (\bibinfo{year}{2016}).

\bibitem[{\citenamefont{Cheng et~al.}(2008)\citenamefont{Cheng, Sheng, and
  Ma}}]{Cheng_PRB08}
\bibinfo{author}{\bibfnamefont{Y.~Q.} \bibnamefont{Cheng}},
  \bibinfo{author}{\bibfnamefont{H.~W.} \bibnamefont{Sheng}}, \bibnamefont{and}
  \bibinfo{author}{\bibfnamefont{E.}~\bibnamefont{Ma}}, \bibinfo{journal}{Phys.
  Rev. B} \textbf{\bibinfo{volume}{78}}, \bibinfo{pages}{014207}
  (\bibinfo{year}{2008}).

\bibitem[{\citenamefont{Soklaski et~al.}(2013)\citenamefont{Soklaski, Nussinov,
  Markow, Kelton, and Yang}}]{Soklaski_PRB13}
\bibinfo{author}{\bibfnamefont{R.}~\bibnamefont{Soklaski}},
  \bibinfo{author}{\bibfnamefont{Z.}~\bibnamefont{Nussinov}},
  \bibinfo{author}{\bibfnamefont{Z.}~\bibnamefont{Markow}},
  \bibinfo{author}{\bibfnamefont{K.~F.} \bibnamefont{Kelton}},
  \bibnamefont{and} \bibinfo{author}{\bibfnamefont{L.}~\bibnamefont{Yang}},
  \bibinfo{journal}{Phys. Rev. B} \textbf{\bibinfo{volume}{87}},
  \bibinfo{pages}{184203} (\bibinfo{year}{2013}).

\bibitem[{\citenamefont{Widmer-Cooper and Harrowell}(2006)}]{Harrowell06}
\bibinfo{author}{\bibfnamefont{A.}~\bibnamefont{Widmer-Cooper}}
  \bibnamefont{and}
  \bibinfo{author}{\bibfnamefont{P.}~\bibnamefont{Harrowell}},
  \bibinfo{journal}{Phys. Rev. Lett.} \textbf{\bibinfo{volume}{96}},
  \bibinfo{pages}{185701(4)} (\bibinfo{year}{2006}).

\bibitem[{\citenamefont{Abe et~al.}(2006)\citenamefont{Abe, Shimono, Ode, and
  Onodera}}]{Abe_AM06}
\bibinfo{author}{\bibfnamefont{T.}~\bibnamefont{Abe}},
  \bibinfo{author}{\bibfnamefont{M.}~\bibnamefont{Shimono}},
  \bibinfo{author}{\bibfnamefont{M.}~\bibnamefont{Ode}}, \bibnamefont{and}
  \bibinfo{author}{\bibfnamefont{H.}~\bibnamefont{Onodera}},
  \bibinfo{journal}{Acta Materialia} \textbf{\bibinfo{volume}{54}},
  \bibinfo{pages}{909 } (\bibinfo{year}{2006}).

\bibitem[{CuZ()}]{CuZr_crystal}
\emph{\bibinfo{title}{$\mbox{CuZr}$ crystal structure: Datasheet from ``pauling
  file multinaries edition -- 2012'' in springermaterials
  (http://materials.springer.com/isp/crystallographic/docs/sd{\_}0450958)}}.

\bibitem[{\citenamefont{Wu et~al.}(2013{\natexlab{b}})\citenamefont{Wu, Li,
  Wang, Song, and Liu}}]{Wu_JCP13}
\bibinfo{author}{\bibfnamefont{Z.~W.} \bibnamefont{Wu}},
  \bibinfo{author}{\bibfnamefont{M.~Z.} \bibnamefont{Li}},
  \bibinfo{author}{\bibfnamefont{W.~H.} \bibnamefont{Wang}},
  \bibinfo{author}{\bibfnamefont{W.~J.} \bibnamefont{Song}}, \bibnamefont{and}
  \bibinfo{author}{\bibfnamefont{K.~X.} \bibnamefont{Liu}},
  \bibinfo{journal}{The Journal of Chemical Physics}
  \textbf{\bibinfo{volume}{138}}, \bibinfo{pages}{074502}
  (\bibinfo{year}{2013}{\natexlab{b}}).

\bibitem[{\citenamefont{Levchenko et~al.}(2011)\citenamefont{Levchenko, Evteev,
  Belova, and Murch}}]{Levchenko_ActaMat11}
\bibinfo{author}{\bibfnamefont{E.~V.} \bibnamefont{Levchenko}},
  \bibinfo{author}{\bibfnamefont{A.~V.} \bibnamefont{Evteev}},
  \bibinfo{author}{\bibfnamefont{I.~V.} \bibnamefont{Belova}},
  \bibnamefont{and} \bibinfo{author}{\bibfnamefont{G.~E.} \bibnamefont{Murch}},
  \bibinfo{journal}{Acta Materialia} \textbf{\bibinfo{volume}{59}},
  \bibinfo{pages}{6412 } (\bibinfo{year}{2011}).

\end{thebibliography}
\end{document}